\documentclass[prl,twocolumn,showpacs,floatfix,superscriptaddress,showpacs]{revtex4}

\usepackage{graphics}
\usepackage{graphicx}
\usepackage{epsfig}
\usepackage{amssymb}
\usepackage{amsmath}
\usepackage{color}

\begin{document}
\def \tr{{\mbox{tr~}}}
\def \ra{{\rightarrow}}
\def \ua{{\uparrow}}
\def \da{{\downarrow}}
\def \be{\begin{equation}}
\def \ee{\end{equation}}
\def \ba{\begin{array}}
\def \ea{\end{array}}
\def \bea{\begin{eqnarray}}
\def \eea{\end{eqnarray}}
\def \nn{\nonumber}
\def \l{\left}
\def \r{\right}
\def \half{{1\over 2}}
\def \etal{{\it {et al}}}
\def \cH{{\cal{H}}}
\def \cM{{\cal{M}}}
\def \cN{{\cal{N}}}
\def \cQ{{\cal Q}}
\def \cI{{\cal I}}
\def \cV{{\cal V}}
\def \cG{{\cal G}}
\def \cF{{\cal F}}
\def \cZ{{\cal Z}}
\def \bS{{\bf S}}
\def \bI{{\bf I}}
\def \bL{{\bf L}}
\def \bG{{\bf G}}
\def \bQ{{\bf Q}}
\def \bR{{\bf R}}
\def \br{{\bf r}}
\def \bu{{\bf u}}
\def \bq{{\bf q}}
\def \bk{{\bf k}}
\def \bz{{\bf z}}
\def \bx{{\bf x}}
\def \bpsi{{\bar{\psi}}}
\def \tJ{{\tilde{J}}}
\def \W{{\Omega}}
\def \lam{{\lambda}}
\def \L{{\Lambda}}
\def \a{{\alpha}}
\def \t{{\theta}}
\def \b{{\beta}}
\def \g{{\gamma}}
\def \D{{\Delta}}
\def \d{{\delta}}
\def \w{{\omega}}
\def \s{{\sigma}}
\def \f{{\varphi}}
\def \x{{\chi}}
\def \eps{{\epsilon}}
\def \h{{\eta}}
\def \G{{\Gamma}}
\def \z{{\zeta}}
\def \hatt{{\hat{\t}}}
\def \hn{{\bar{n}}}
\def \vk{{\bf{k}}}
\def \vq{{\bf{q}}}
\def \gk{{\g_{\vk}}}
\def \nd{{^{\vphantom{\dagger}}}}
\def \yd{^\dagger}
\def \av#1{{\langle#1\rangle}}
\newcommand{\ket}[1]{|{#1}\rangle}
\newcommand{\bra}[1]{\langle{#1}|}
\newcommand{\braket}[1]{\langle{#1}\rangle}
\newcommand{\ad}{a^\dagger}
\newcommand{\e}{\ensuremath{\mathrm{e}}}
\newcommand{\norm}[1]{\ensuremath{| #1 |}}
\newcommand{\aver}[1]{\ensuremath{\big<#1 \big>}}

\newcommand{\cred}{\color{red}}

\title{Quench dynamics and non equilibrium phase diagram of the Bose-Hubbard model}

\author{Corinna Kollath}
\affiliation{Universit\'{e} de Gen\`{e}ve, 24 Quai Ernest-Ansermet, CH-1211 Gen\`{e}ve, Switzerland}
\author{Andreas M. L\"auchli}
\affiliation{Institut Romand de Recherche Num\'erique en Physique des Mat\'eriaux (IRRMA), CH-1015 Lausanne, Switzerland}
\author{Ehud Altman}
\affiliation{Department of Condensed Matter Physics, The Weizmann Institute
of Science, Rehovot 76100, Israel}

\date{\today}
\begin{abstract}
We investigate the time evolution of correlations in the Bose-Hubbard model
following a quench from the superfluid to the Mott insulator.
For large values of the final
interaction strength the system approaches a distinctly
non-equilibrium steady state that bears strong memory of the initial conditions.
In contrast, when the final interaction strength is comparable to the hopping,
the correlations are rather well approximated by those
at thermal equilibrium. The existence of two distinct non-equilibrium
regimes is surprising given the non-integrability of the Bose-Hubbard model.
We relate this phenomenon 
 to the role of quasi-particle interactions in the Mott insulator.
\end{abstract}
\pacs{
03.75.Lm       
05.70.Ln 
67.40.Fd
73.43.Nq   
}

\maketitle

Recent experiments with ultra cold atomic gases have opened
exciting possibilities for studying non-equilibrium quantum
dynamics of many body systems. In particular, the
high degree of tunability
allows one to rapidly change system parameters
and observe the subsequent quantum evolution. Furthermore,
thanks to the almost perfect
isolation of the atoms from the environment, the quantum dynamics
can remain coherent for exceedingly long times.
These advantages were used, for example, to study non adiabatic
dynamics across the quantum phase
transition between a superfluid and a Mott insulator
\cite{GreinerBloch2002, GreinerBloch2002b}, as well as
the crossover of paired fermion superfluids from
weak to strong coupling \cite{GreinerJin2005,ZwierleinKetterle2005}.

In many cases the system parameters are changed so fast, that 
one may consider the sudden limit:
The system is prepared in the ground state of an initial Hamiltonian $H_i$, and
then evolves under the influence of a different Hamiltonian $H_f$.
Fundamental questions that arise 
concern the approach of the system to a new steady state and
the nature of this steady state.
Does it retain memory of the initial state? How is
it related to the thermal equilibrium of $H_f$?
These questions were addressed in a number of recent works, by solving
various integrable models~\cite{IgloiRieger2000,SenguptaSachdev2004,CherngLevitov2005, CalabreseCardy2006,RigolOlshanii2006,Cazalilla2006}.
In these systems the long time steady state was found to be non thermal and
often carried memory of the initial state. In a fascinating experiment, Kinoshita
{\it et al.} \cite{KinoshitaWeiss2006} investigated the thermalization of strongly interacting
ultra cold atoms in a {\em nearly} integrable situation. The result, at the maximal
time scale of the experiment, was a non thermalized steady state.

In this Letter we numerically investigate the evolution of correlations following a sudden
change of parameters in the {\em non-integrable} Bose Hubbard model (BHM)~\cite{FisherFisher1989},
describing 
cold atomic gases in optical lattices \cite{JakschZoller1998, Zwerger2003}
\begin{equation}
\label{eq:bh}
H= -J \sum_{\langle i,j\rangle} \left(b_i^\dagger b^{\phantom{\dagger}}_{j}+h.c.\right)
+ \frac{U}{2} \sum _{i} n_i ( n_i-1).
\end{equation}
In equilibrium at integer filling, this model exhibits a quantum
phase transition at a critical value of the interaction strength
$U/J=u_c$, between a superfluid ($U/J<u_c$) and a Mott insulating
state ($U/J>u_c$)~\cite{FisherFisher1989}. In a one-dimensional
system with unit filling $u^{1D}_c\approx 3.37 $~\cite{KuehnerMonien2000} and in two dimensions
$u^{2D}_c\approx  16.7$~\cite{ElstnerMonien1999}.

\begin{figure} [t]
\centerline{\includegraphics[width=\linewidth]{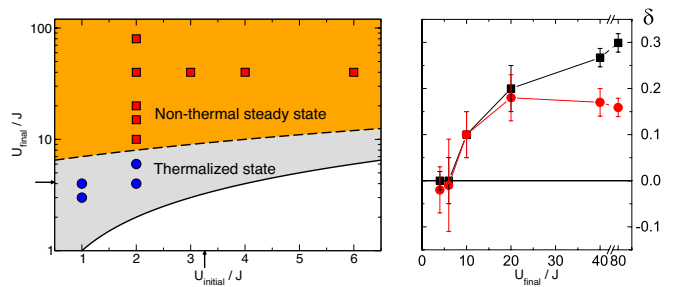}}
\caption{(Color online) Left panel: Non equilibrium phase diagram in
the space of initial and final interaction strengths. Two
regions are found, one where the steady state is distinctly non thermal and
the other where correlations do appear to thermalize within the numerical
error bounds. Squares and
circles mark points in the respective regions where numerical
results for a one-dimensional system were obtained. The full line 
follows $U_i=U_f$. 
Small arrows mark the equilibrium critical value for an infinite one-dimensional system.
Right panel: Difference between the correlation
$\av{b^\dagger_i b_{i+r}}$ at steady state and its nominal value for
the ground state of the final Hamiltonian. 
Data is shown for $r=1$ (circles) and $r=2$ (squares) along the cut
$U_i/J=2$. The crossover from slower than ground state decay
($\delta>0$) to faster than ground state is seen at $U_f\approx 6J$.
\label{fig:phasediagram}}
\end{figure}

Our study is motivated by the experiment with ultra cold bosonic atoms in
an optical lattice \cite{GreinerBloch2002b}, where
the lattice intensity was increased suddenly, taking the system from
the superfluid phase into the Mott insulator regime.
Following the quench, a remarkable series of
collapse and revivals of the interference pattern was observed,
which relaxed after a few oscillations. What processes are responsible for the relaxation
and what is the nature of the steady state that is reached?
The general expectation for non-integrable models like (\ref{eq:bh})
is that the long time steady state will be essentially equivalent to
thermal equilibrium. Surprisingly, we find that this is not always the case.
When the interaction $U_f$ in the final state is much larger than $J$
the system reaches
a quasi-steady state that is very different from thermal equilibrium and
retains memory of the initial state. As $U_f$ decreases
the nature of the steady state changes, and in the region where $U_f$ is comparable
to $J$ the steady state correlations are well approximated by those
at thermal equilibrium (cf.~Fig.~\ref{fig:phasediagram}). We relate this crossover in the non-equilibrium
behavior to the ineffectiveness of quasi-particle interactions in the strongly
interacting regime.

The revivals seen in the interference patterns in the experiment \cite{GreinerBloch2002b} are
easy to understand in the limit $J\to 0$.
In this limit the evolution
operator is site factorizable and given by  $ \prod_i\e^{\textrm{i} U_f
  n_i(n_i-1)t/2}$. This implies periodic time dependence, since the operator $
n_i(n_i-1)/2$ takes integer values for any Fock state. An arbitrary initial wave
function 
therefore revives entirely after times $\tau_n=2\pi n/U_f$ where $n$ is integer
($\hbar$ is set to unity throughout). A non vanishing hopping matrix element
$J$ greatly complicates the time evolution, and leads to a relaxation of
the oscillations.
In the following we investigate this situation 
using numerical methods and then interpret the results within a tractable effective model.
We implement the dynamic transition from the superfluid to the Mott-insulating regime
at unit filling ($n=1$) by a sudden quench of the interaction $U$ at fixed hopping $J$. Thus the initial "superfluid" wavefunction is the ground
state of the Hamiltonian $H_i$ with $U=U_i<U_c$, and it evolves subject to the Hamiltonian $H_f$
with $U_f$ in the Mott insulator regime or close to the critical point.
The time evolution of the many-body wavefunction is computed by exact diagonalization (ED) based on Lanczos-type
methods~\cite{park:5870,manmana-2005-789} and by the adaptive time-dependent density matrix renormalization
group method (adaptive t-DMRG)~\cite{WhiteFeiguin2004, DaleyVidal2004,GobertSchuetz2004}. The ED is used to
study 1D and 2D systems with up to 18 sites, while the adaptive t-DMRG
is used for 1D systems with up to 64 sites keeping up to 200 DMRG
states.
For computational reasons the Hilbert space on each site was truncated at high occupation numbers. In ED (t-DMRG) we typically kept up to 4 (9) bosons per site.

\begin{figure}
  \centerline{\includegraphics*[width=0.98\linewidth]{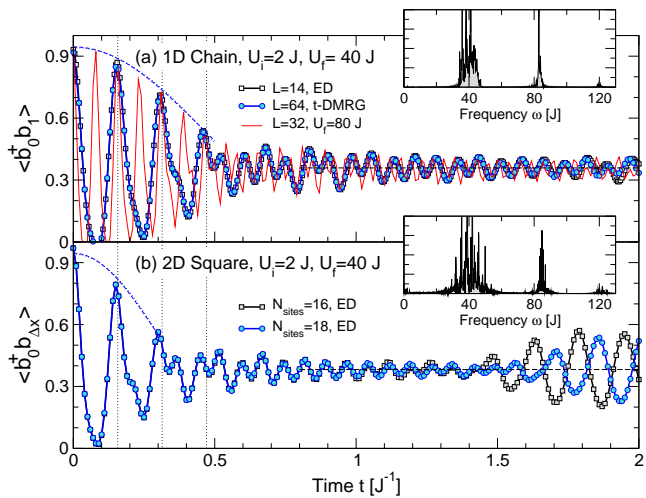}}
\caption{(Color online) Short time behaviour of the nearest-neighbor correlation functions for $U_i=2J$ and $U_f=40J$.
  (a) t-DMRG and ED results for 1D chains. For comparison the thin full line without symbols shows a
  relaxation for $U_f=80J$.
  (b) ED results for 2D square lattices.
  In both cases there are clear oscillations with a period $2\pi/U_f$ (dotted vertical lines for $U_f=40J$),
  which relax on a time scale of $1/J$. Revival
  of the correlations at $\approx 1.5/J$ in (b) is due to finite size effects.
  The two insets show the Fourier transform of the oscillations. The main weight lies in a broad
  band at $\omega \approx U_f$, with some smaller bands at higher multiples of
  $U_f$.
  \label{fig:shorttime}}
\end{figure}
For a sudden quench deep into the Mott-insulating regime
we find that a partial revival of the wave-function survives. This can for
example be seen in the off-diagonal correlation functions
$\av{b\yd(x,t)b\nd(0,t)}$ which display oscillations
with a period set by $2\pi/U_f$. 
The oscillations relax to a
quasi-steady state on a time scale $\sim 1/J$.
We find that this relaxation time, as well as
the value of the correlation reached in the quasi-steady state, are independent of $U_f$ for sufficiently
large $U_f$. Fig.~\ref{fig:shorttime} shows an example of the evolution of nearest neighbor
correlations. A similar evolution is observed for longer range correlations.
Later we use a simple model valid at strong coupling to show that the relaxation time
is related to the existence of a quasi-particle band of width $\sim J$ around an energy $U_f$.
The short range correlations oscillate with all the frequencies in this band,
and therefore they dephase after a time scale of the order of the inverse band width.
Indeed a Fourier decomposition of the oscillations (inset of Fig.~\ref{fig:shorttime})
reveals this band, as well as weaker contributions from higher multiples of
$U_f$. The amplitude of the oscillations with frequencies of higher multiples
depends strongly on the particle distribution of the initial state.
\begin{figure}
  \centerline{\includegraphics*[width=0.97\linewidth]{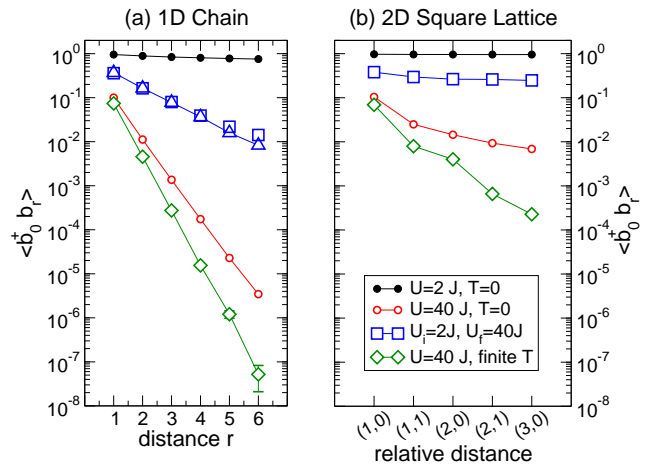}}
  \caption{(Color online) Decay of the correlations $\av{b^\dagger_i b_{ i+r}}$ with distance $r$
  after a quench from $U_i=2J$ to $U_f=40J$.
  Squares (ED) and triangles (t-DMRG) show the averaged results of the
  quasi steady state. The average value is
  determined fitting a linear function to the results between
  $t_1\approx J^{-1}$ and $t_2=20J^{-1}$.
  Diamonds show equilibrium QMC results at finite temperature
  (see text for details), and the filled and open circles display the $T=0$ correlations in the ground state for
  $U_i$ and $U_f$, respectively.
  \label{fig:ltcorr}}
\end{figure}
\begin{figure}
 \centerline{\includegraphics*[width=0.9\linewidth]{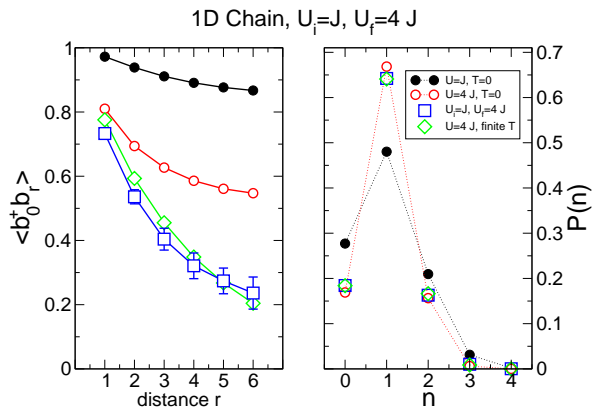}}
 \caption{(Color online)
 Left panel: Decay of the correlations $\av{b^\dagger_i b_{i+r}}$ with distance $r$
 after a quench from $U_i=J$ to $U_f=4 J$.
 Squares show the averaged results for times between $t_1\approx J^{-1}$ and
 $t_2=20 J^{-1}$ determined as in Fig.~\ref{fig:ltcorr}. 
 Diamonds show equilibrium
 QMC results at finite temperature ($T\approx 0.8 J$) and $U=4J$, and the filled and open circles
 display the $T=0$ correlations in the ground state for $U_i$ and $U_f$, respectively.
 Right panel: Particle number distribution $P(n)$. The labeling is the same as in the left panel.
  \label{fig:corr_1_4}}
\end{figure}

We turn to investigate the nature of the quasi-steady state reached by the off-diagonal
correlations. The general expectation
in non-integrable systems, is that the correlations relax to thermal equilibrium.
The temperature at this equilibrium is set by the internal energy imposed on the system
by the initial conditions. Interestingly, in spite of the non-integrability
of the BHM we find two different regimes of behavior
depending mostly on the magnitude of the interaction strength in the final state.
When $U_f$ is very large the correlations in the steady state bear strong memory
of the initial state. In particular their decay with distance is much slower than the corresponding thermal
correlations and even slower than the ground state correlations at the final point.
This behavior is shown in Fig.~\ref{fig:ltcorr}. The equilibrium
finite temperature correlations were calculated using quantum Monte Carlo (QMC)
simulations of the BHM (\ref{eq:bh}) with $U=U_f$, where the temperature was
determined by matching the on-site particle distribution $P(n)$.
This yielded $T_\mathrm{1D}\approx 21.5J$ and $T_\mathrm{2D} \approx 23.3 J$.
A completely different regime is realized when the final state is closer to
the superfluid transition, i.e. $U_f\lesssim 6J$.
In that case the correlations at long times do decay with distance
faster than the ground state correlations and
fit reasonably well to correlations at thermal equilibrium as calculated using
QMC. The good fit of the nearest-neighbor correlations implies together with the
  matching of the particle distribution that the temperature used for
  the QMC simulations corresponds to the energy
forced into the system by the quench. An example of the correlations in this regime is displayed in the
left panel of Fig.~\ref{fig:corr_1_4}.
Contrary to the regime of large $U_f$, the oscillations
with period $2\pi /U_f $ are overdamped. We note however that
in this regime the correlations do not reach a true steady state
within the time-scale we are able to simulate reliably.

In contrast to the rich behavior of the off diagonal correlations,
on-site quantities such as the particle distribution $P(n)$ reach a
much simpler steady state for all $U_f$ we considered.
For large $U_f$, $P(n)$ almost does not relax from the values in the initial
state, in agreement with mean-field calculation for the case of high fillings
\cite{SchuetzholdFischer2006}.
For smaller $U_f$ some relaxation occurs, see right panel of Fig.~\ref{fig:corr_1_4}.
In all cases we were able to determine a temperature where the QMC simulations
at $U_f$ reproduce the long-time behavior of $P(n)$ to good accuracy.

The essential results of the calculations in different parameter
regimes are summarized in Fig.~\ref{fig:phasediagram}. The left
panel presents a non-equilibrium phase diagram in the plane of
$U_i/J$ and $U_f/J$. In one region (large $U_f$) off-diagonal
correlations reach a distinctly non thermal steady state. In the
other region, the correlations after some time, are well described
by the thermal equilibrium results. A cut 
along the
line $U_i/J=2$ shows a clear crossover of the steady state
correlations from decaying slower than ground state correlations at
$U_f$ to faster than ground state (as expected of thermal
correlations) at $U_f\approx 6J$.

We now argue that the different steady state regimes may be understood
qualitatively based on a simple model \cite{AltmanAuerbach2002}.
One can think of the excitations of a Mott insulator
as particles and holes ($p\yd_i$ and $h\yd_i$)
that hop on the lattice and may be created and annihilated in pairs.
From this point of view, the initial state (a commensurate superfluid)
is a simultaneous condensate of particles and holes.
Its time dependence is determined by the effective
Hamiltonian in the Mott regime. Neglecting quasi-particle
interactions, this Hamiltonian is diagonalized by
a Bogoliubov transformation, so that $H_0=\sum_{\bk,\a}\w_\bk \b\yd_{\a\bk} \b\nd_{\a\bk}$,
where $\b\yd_{\a\bk}$ creates a quasi-particle of the Mott insulator.

Now consider the time evolution of the momentum distribution
$\av{n_\bk}=\av{b\yd_\bk b\nd_\bk}$. In terms of the particles and
holes, a boson is given roughly by the combination
$b\yd_\bk\sim p\yd_\bk+ h\nd_{-\bk}$.
Therefore the component
$\av{n_\bk(t)}$ of the momentum distribution can be constructed from
quasi-particles of the Mott insulator carrying momenta $\bk$ and $-\bk$,
and it oscillates at a frequency $\omega_\bk$.
This fact bears on the dynamics of spatial
correlations $\av{b\yd_{j+r}b\nd_j}$, which are
the Fourier transform of $\av{n_\bk}$. Short range correlations receive
contributions from all $\bk$ components, and therefore they
dephase at a rate comparable to the full quasi-particle bandwidth.
Long range correlations on the other hand are dominated by $n_\bk$
at small $k$.

Within this picture thermalization occurs due to quasi-particle interactions.
In particular the quasi-particle population equilibrates because of quartic
processes of the form $\b\yd_\bq\b\nd_{\bk+\bq/2}\b\nd_{-\bk+\bq/2}\b\nd_0$.
Note that quartic terms that conserve the quasi-particle number cannot
by themselves induce thermalization because they do not change the non-equilibrium
quasi-particle population forced on the system by the initial conditions.
At the level of Fermi's golden rule, the process mentioned above may occur only while conserving energy.
But this is impossible deep in the Mott insulator, as long as the gap $\D$
is larger than half the quasi-particle bandwidth $W\sim 4z J$. We conclude that thermalization
should be effectively suppressed in the regime $U\gg J$, in agreement with the
numerical results. Of course there are higher order processes that may still induce
thermalization, but apparently this occurs at time scales much larger than the relaxation
time $1/J$. On the other hand, as the final state approaches the critical point $\D=0$,
quasi-particle interactions become increasingly effective, which leads to
rapid thermalization. The simple picture outlined above is strictly valid
only for a dilute quasi-particle population, that is for initial state close
to the transition. However from the numerical results it seems
to apply more generally.

The results of the present study
leave open interesting conceptual questions.
Is the non thermal state reached a true steady state?
Is there a longer time scale associated with reaching
true equilibrium as in the prethermalized states discussed
in Ref.~\cite{BergesWetterich2004}, or a scale free relaxation
analogous to aging phenomena?

We propose that some of these questions may be addressed by experiments
with ultra cold atoms on optical lattices. The simplest observable
to consider is the visibility of the interference pattern as
defined in \cite{GreinerBloch2002b}.
For example, in a one dimensional homogenous tube, following a quench from
$U_i=2J$ to $U_f=40J$ we expect, based on t-DMRG calculation, that
the visibility will relax to a value of approximately $60\%$, compared
to visibility of only $20\%$ in the ground state of the system
with $U=U_f$.

An obstacle for the interpretation of experiments is the
existence of additional sources of relaxation, most prominently
the confining potential and the presence of many parallel tubes.
 We performed time dependent Gutzwiller
calculations to
compare the effect of the confining potential to that of
tunneling. The effect of the tunneling dominates, if the
energy difference between neigbouring site $V_0 (2j-1)$ is smaller than the
width of the particle-hole energy bands $\sim 6zJ$, i.e. $ V_0 (2j_m-1)/(6zJ) \lesssim 1$. Here $V_0$ is the
prefactor of the trapping potential and $j_m$ is the extension of the condensate.
Substituting into this analysis the experimental parameters of \cite{GreinerBloch2002b} it follows that the main source
of dephasing in that experiment was the confining potential.
However it is not difficult to reach the regime where the tunneling
gives the dominant contribution.


\acknowledgments
We thank B.~Altshuler, E.~Demler, T.~Giamarchi, S.~Kehrein, S.~Manmana,
A.~Muramatsu, R.~Noack, F.~Werner, and S.~Wessel for fruitful discussions. This
work was partly supported by the SNF under MaNEP and
Division II, and by the DFG grant HO 2407/2-1. EA acknowleges support
of the U.S.-Israel binational science foundation.
We acknowledge the allocation of computing time at CSCS (Manno) and LRZ (Garching).
The QMC simulations have been performed using the SSE application \cite{SandvikSSE1999,ALPS-SSE}
of the ALPS project \cite{ALPS-Main,ALPS-Scheduler}.

\end{document}